\documentclass[aps,prl,twocolumn,groupedaddress,showpacs]{revtex4}
\usepackage{graphicx}
\usepackage{isolatin1}
\newcommand{\myscalebox}[1]{\scalebox{0.42}[0.42]{#1}}
\begin{document}
\title{ Corrections to Scaling are Large for Droplets in
  Two-Dimensional Spin Glasses }
\author{A. K. Hartmann}
\affiliation{Institut f\"ur Theoretische Physik, Universit\"at G\"ottingen,
Bunsenstra\ss{}e 9, 37037 G\"{o}ttingen, Germany}
\author{M. A. Moore}
\affiliation{Department of Physics and Astronomy, University of 
Manchester,
Manchester, M13 9PL, United Kingdom} 
\date{\today}
\begin{abstract}
The energy of a droplet of linear extent $l$ in the droplet theory
of spin glasses goes as $l^{\theta}$ for large $l$. It is shown  by
numerical studies of large droplets in two dimensional systems that this 
formula needs to be modified by the addition of a scaling correction 
$l^{-\omega}$ in order to accurately describe droplet energies at the length
scales currently probed in numerical simulations. Using this simple
modification it is now possible to explain many results which have been found
in simulations of {\em three} dimensional Ising spin glasses with the droplet
model.
\end{abstract}
\maketitle
There are two rival theories to describe the ordered state
of spin glasses: the droplet theory 
\cite{McM,BM,FH} and the replica symmetry breaking (RSB) theory of Parisi
\cite{Parisi,sgb}. The droplet picture assumes that the
low-temperature behavior is  
governed by droplet-like excitations, where
 excitations of linear spatial extent $l$ typically cost an energy
of order $l^{\theta}$. Thus in
the thermodynamic limit the excitations which flip a finite fraction of the
spins cost an infinite amount of energy if $\theta>0$. These excitations are
expected to be compact and their surface 
has a fractal dimension $d_s<d$, where $d$ is the space dimension.
  Furthermore it is usually
assumed that the energy  of different types of excitations,
e.g. droplets and domain walls, induced by changing the boundary
conditions, are described by the same exponent $\theta$. On the other hand
 the RSB theory assumes 
that there are low-energy droplet excitations in which a 
finite fraction of the 
spins in the system are reversed but which only cost a finite amount of energy
in the thermodynamic limit. The surface of these excitations is space-filling 
so that the fractal dimension of their surface $d_s=d$ \cite{Marinari}.

Many numerical studies have been done in an attempt
to resolve the controversy, but studies in three \cite{3D} and four
 dimensional systems \cite{4D}
are limited in the range of systems sizes which can be studied.
In two dimensions, however, much larger systems can be studied. For this
 dimension, no spin-glass
order exists at any finite temperature \cite{SG-kawashima1997}, 
but there are still puzzling results concerning droplet excitation energies 
about the groundstate. 
The energies of cross-system domain walls \cite{perturb, HY} and droplet-like
excitations \cite{KA,picco2001} are apparently descibed by
 different exponents, which is not compatible with the (usual) droplet
picture and might be regarded as a vestige of RSB type behavior.

In this letter we
study numerically droplets in two-dimensional spin
glasses similar to the droplets studied in Ref. \onlinecite{KA}, but up 
to even
larger sizes. We show that the
numerical results mentioned above can be explained by a
simple scaling correction.
For larger droplets we find a behavior which is compatible with the
droplet prediction, i.e. the exponent $\theta$ is the same for
droplets and cross-system domain-walls. Smaller droplets appear to be
described by an effective exponent $\theta^{'}$. We also study how the
volume of an average droplet behaves with system size. We find that
when considering just smaller system sizes, droplets appear non-compact,
but for larger systems,  crossover to a
compact behavior looks likely. Thus in two-dimensions
the droplet picture seems to be entirely valid.  
Finally, we discuss also some recent
numerical results \cite{LQ, KM, PY} in three dimensions, which also may be 
explained
within the droplet picture using the correction to scaling for droplet energies
introduced here.

We consider 
the Hamiltonian which is usually studied in numerical work of the Ising
spin glass model:
\begin{equation}
{\mathcal H}=-\sum_{\langle i,j\rangle }J_{ij}S_iS_j,
\end{equation}
where the sites $i$ lie on the sites of a cubic lattice with $N=L^d$ sites,
$S_i=\pm1$, the $J_{ij}$ have a Gaussian distribution of zero mean and
unit variance and couple nearest-neighbor sites on the lattice.

It is important to realize what the low-energy excitations look like in spin
glasses. Consider a domain wall crossing a system  $M\times M\times L$. The
energy cost of this domain on the droplet picture will be of order
$M^{\theta}$ and the domain wall will be fractal with dimension $d_s$ and 
with an area of order $M^{d_s}$. In three or more dimensions the 
interface may have holes through it. 
Because it is fractal the extent of this wandering
is of order $M$. The wandering of the interface by an amount of order $M$ 
affects the determination of $\theta$, see Ref. \onlinecite{Carter},
where it was found that the best results were obtained when $L\gg M$ as then
the interface is not affected by its interactions with the ends of the system.
Suppose now we have two domain walls across the system. 
Then if their separation 
is large compared to $M$ they will be unaffected by the presence of the other.
However, if they are closer together they interfere with each other and
their overall energy will be greater than if the other one were absent. This
is because domain walls wander to take advantage of weak bonds and 
since the domain walls cannot cross, the presence of the second domain wall
will preclude the first domain wall from cutting through bonds it would
have cut in the absence of the second wall. As a consequence of this,
domain walls effectively repel each other. If they have a 
separation of order $l$, then the repulsive energy between them would
be expected to vary with their separation as a power law,
$l^{-\omega^{'}}$. No investigations of $\omega^{'}$ seem to exist in
the literature.
 
Consider now a droplet of linear extent $l$. It too will have a fractal
surface described by $d_s$. Pictures of large droplets have now appeared 
\cite{KM}, and a systematic investigation of them is in 
\cite{HKM}. Their fractal nature ensures that in three or more
dimensions that they have holes through them,
giving them a sponge-like appearance. However, because the surface of the
droplet may wander by a distance of order $l$ the energy of the droplet $E$
will be modified by its wandering and ``collisions'' with itself to a form 
which we suppose by analogy with the above is 
\begin{equation}
E=Al^{\theta}+Bl^{-\omega}.
\label{basic}
\end{equation}
The term $Bl^{-\omega}$ is a scaling correction to the form of the droplet
energy at large $l$ and $A$ and $B$ are positive constants. We will
show that the scaling correction is very significant in two
dimensions.

In this
dimension $\theta$ is very accurately determined as systems of $480^2$
can be studied \cite{HY}. The
exponent associated with a single domain wall, that is $\theta$, is 
approximately
 $-0.29$ \cite{Carter,HY}. However, studies where, say, $\theta$ is 
determined from
the effects of thermally excited droplets (such as Monte Carlo simulations of
the spin-glass susceptibility)  yield an apparent $\theta^{'}$ close
to $-0.47$ \cite{KA,picco2001}. 
This discrepancy has long been a puzzle, and has 
prompted suggestions that perhaps different exponents describe
domain wall energies and droplet energies \cite{KA}. However, Eq. (\ref{basic})
offers a way of resolving the
discrepancy. For small values of $l$, droplets would apparently have 
energies decreasing as $l^{-|\theta^{'}|}$ when their energies are affected
by the correction to scaling term, but at large values of $l$ the decrease
with $l$ will be slower and be as $l^{-|\theta|}$, as expected from 
the conventional droplet approach. 

We have tested this explicitly by studying 
larger sizes than before using an exact ground-state (GS)
algorithm \cite{opt-phys2001}.
For the special case of a planar system without 
magnetic field, e.g. a square lattice with periodic boundary conditions 
in at most one direction, there are efficient polynomial-time ``matching'' 
algorithms \cite{bieche1980}. The basic idea is to represent each realization 
of the disorder by its frustrated plaquettes \cite{toulouse1977}. Pairs of
frustrated plaquettes are connected by paths in the lattice and the weight 
of a path is defined by the sum of the absolute values of the coupling 
constants which are crossed by the path. A ground state corresponds to the 
set of paths with minimum total weight, such that each frustrated plaquette 
is connected to exactly one other frustrated plaquette. This is called a 
minimum-weight perfect matching. The bonds which are crossed by paths
connecting the frustrated plaquettes are unsatisfied in the ground state, 
and all other bonds are satisfied. 


We want to mimic the droplet generation
used in Ref. \onlinecite{KA}, where a center spin was forced to reverse its
orientation with respect to its orientation in the  GS,
the boundary spins were fixed, and a new
GS under these constraints was calculated. Note that fixing the absolute
orientation of a spin does not allow for a fast polynomial algorithm.
Hence, a different approach must be used to study large systems.

Our generation of the droplets works in the following way. For each 
realization, first a ground state is calculated with free boundary
conditions in both directions. Then several \lq\lq hard" bonds are introduced,
i.e. bonds
with a high value of the absolute strength (e.g. 
$|J_{ij}|=2N\times \max_{\langle i,j \rangle}\{J_{ij}\}$). 
If the subsystem of hard
bonds does not exhibit frustration, no hard bond will be broken, if a new
GS is calculated. First, all boundary spins are fixed relative to each
other by hard bonds around the border 
chosen to be compatible with their GS orientations, 
i.e. 
the bonds between pairs of boundary spins are replaced. Second, a line of hard
bonds is created which runs from  the middle of (say) the left border
to a pre-chosen center spin, 
again fixing the pair's spins in their relative GS orientations. 
Next, the sign of exactly one hard bond on this line is inverted. Finally,
a GS of the modified realization is calculated. With respect to
the original GS, the result is a minimum energy 
droplet fulfilling the constraints that it contains the center spin,
does not run beyond the boundary and that it has a surface which runs through
the hard bond which has been inverted. The energy of the droplet is 
defined as the
energy of the resulting configuration calculated using the original bond
configuration.
For each realization, this procedure is iterated over all the bonds which 
are located on the line from the boundary to the center, when in each case
exactly one hard bond is inverted. Among all generated droplets, the
one exhibiting the lowest energy is selected.

This procedure  is similar, but not equivalent to the
method used  in Ref. \onlinecite{KA}. The difference is that the surface 
of our 
droplets crosses the line from the boundary to the center spin exactly
once. This means, it cannot cross this line more than once, i.e. it
cannot \lq\lq meander" across this line. But it may meander in all other
directions. To reduce the influence of this constraint, we have iterated
over all four lines of bonds running from the left, right, top and
bottom boundary to the center spin, and selected the minimum energy
droplet among all the $2L-2$ droplets generated in this way. 
Hence, we have obtained
droplets which may meander in any direction, but not in all four
directions at the same time. 

We have tested explicitly the influence of this constraint by measuring
in how many directions each of the lowest energy droplets meanders. For small
systems meandering does not occur, but meandering increases with system
size. For the largest systems we have considered, about half the
droplets do not meander at all, about 33 \% meander in one direction,
14 \% meander in two directions and three percent in three directions. Hence,
for the system sizes studied, four-way meandering droplets probably 
would occur rarely, if it were allowed by the method. 
Hence, we
believe that our results are not strongly influenced by the constraint
imposed by our algorithm. Furthermore, we have also tested droplets
generated in a different way, which allow for full meandering, but this
approach is restricted to small sizes, see below.

\begin{figure}[htb]
\begin{center}
\myscalebox{\includegraphics{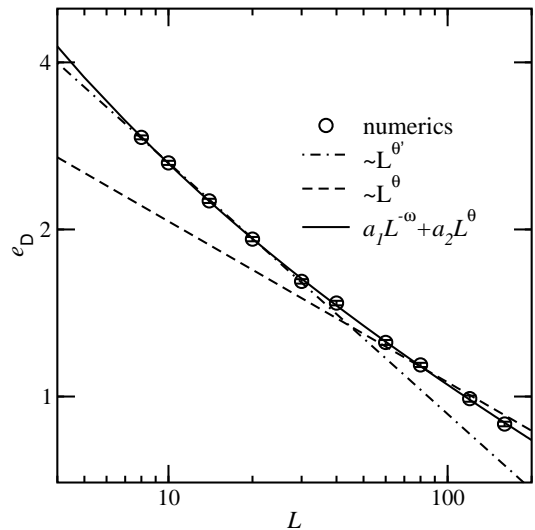}}
\end{center}
\caption{Average droplet  energy $E$ as a function of the system size. The
dashed straight lines represent algebraic functions with exponents 
$\theta^{\prime}=-0.47$ and $\theta=-0.29$, while the full
curved line represents the function from Eq. (\ref{basic}) 
with $\omega=1$, $\theta=-0.29$). }
\label{figEdCross}
\end{figure}

We have generated droplets for system sizes in the range $6\le L \le 160$,
each time 5000 independent realizations.
The average droplet energy as a function of system size is shown in Fig. 
\ref{figEdCross}. For small sizes up to $L=40$, the energy can be
fitted by an algebraic function $\sim L^{\theta^{\prime}}$
with $\theta^{\prime}=-0.45(1)$, which is compatible with the earlier
result \cite{KA} which had been interpreted in favor of the droplet exponent
being different from the domain-wall exponent. 
For larger sizes, a clear crossover can be observed, the 
data is compatible with an exponent $\theta=-0.29$. A fit of the
exponent in $[80,160]$ yields $\theta=-0.32(1)$, which indicates
 that the
thermodynamic limit may not been fully reached. We have also tested
the scaling ansatz Eq. (\ref{basic}) by fixing $\theta=-0.29$, resulting
in $\omega=0.97(5)$ and a very good quality \cite{goodness} $Q=0.81$ 
of the fit.
Curiously, the value for $\omega$ is close to the value which
$\theta$ takes in one dimension,  $-1$.

To estimate the influence of the slightly restricted meandering,
we have also studied droplets, where in addition to the hard bonds
around the boundary, only one additional inverted hard bond was introduced.
For each system, this inverted hard bond was iterated over  {\em all}
bonds connecting at least one non-boundary spin. The minimum energy droplet
was selected among all droplets containing the pre-selected central spin.
Due to the large numerical effort, only sizes up to $L=30$ could be
studied. When considering up to 1000 realizations, the data was
compatible with an algebraic decay with $\theta^{\prime}=-0.47(1)$.
We have increased the statistics, such that we have treated 80000
realizations for small sizes down to 20000 realizations for $L=30$.
Now the straight algebraic fit has a very low quality ($Q=10^{-26}$) compared
to the fit including the correction to scaling
($\omega=-2,\theta=-0.29,Q=0.61$).

\begin{figure}[htb]
\begin{center}
\myscalebox{\includegraphics{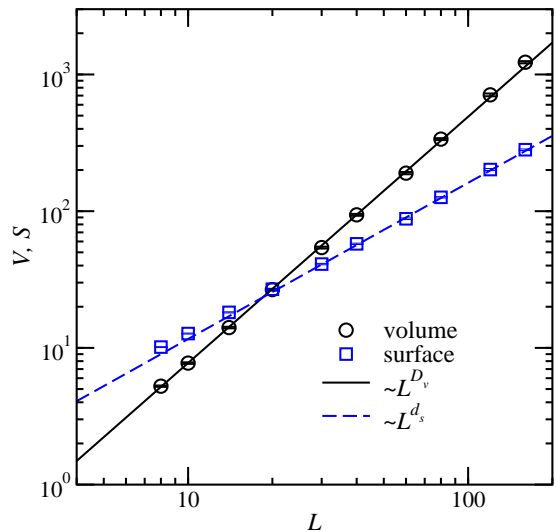}}
\end{center}
\caption{Average droplet volume $V$ and average droplet surface $S$
as a function of the system size. The
straight lines represent algebraic functions with exponents 
$D_v=1.8$ respectively $d_s=1.1$.}
\label{figVS2cross}
\end{figure}

For further comparison with Ref. \onlinecite{KA}, we show in Fig. 
\ref{figVS2cross} the average droplet volume and surface as a function
of the system size. By fitting to algebraic laws, we find a
volume fractal dimension of $D_v=1.80(1)$ and a surface fractal dimension
of $d_s=1.10(1)$, which are both compatible with the results obtained
before. Hence, the crossover observed in the energy is not obvious
when considering volume and surface and the droplets seem to be non-compact
(i.e. $D_v<2$). We have tested, whether this result may be a consequence
of a correction to scaling, by fitting $V(L)$ to a function of the
form $EL^2+FL^{D_v^{\prime}}$ resulting in $D_v^\prime=1.52(5)$ and
a much better quality of the fit; $Q=0.59$ as
opposed to $Q=0.002$ for the
pure algebraic fit. 
Hence a crossover to compact droplets seems
likely. In the figure, both functions would be hard to distinguish,
hence we have omitted including the correction to scaling fit.
We have also
tried a correction to scaling for the surface, since $d_s$ around 1.3
in other studies \cite{perturb}, but using the correction to scaling we
could change $d_s$ in a wide range without affecting the quality of the fit.

A crossover between  large and small $L$ dependencies
was seen in the work of 
Middleton \cite{M}. He studied the scaling of ground-state
link overlaps and observed a gradual evolution of the finite-size
behavior. 
Middleton himself attributed this gradual evolution to averages over 
droplets of all sizes up to the system size $L$. 
This kind of behavior is an additional mechanism, independent of our
 scaling correction for the droplet energies themselves, 
to explain why certain averages   
might be slow to converge to the expected  behavior for large $L$.

We have seen that in two dimensions that the corrections to scaling are
large, so that for medium-size systems they can mask
the true asymptotic behavior. A similar effect would be expected to
occur in higher dimensions and indeed might already have been observed by
Lamarcq et al. \cite{LQ} in their study of  droplets in three dimensions.
They first computed the ground state for both $N=6^3$ and $N=10^3$ spins.
They then chose an arbitrary reference spin and flipped it along
with a cluster containing $V-1$ other spins connected to it. They
next minimized the energy of this cluster by exchange Monte Carlo, but with 
the constraint that the reference spin is held fixed and the cluster was 
always connected and of size $V$. They found the largest extension (mean 
end-to-end distance) $l$ of the 
cluster and found that for $V\leq 33$ its energy $E(l)$ varied as
$l^{\theta^{'}}$, where the exponent $\theta^{'}$ was $-0.13 \pm0.02$. For 
larger
$V$ values they found that $E(l)$ was increasing rather than decreasing with
$l$ but they were unsure whether this might not be an artifact of insufficient
numbers of Monte Carlo steps. Note that according to Eq. (\ref{basic}), 
when both
$\theta$ and $\omega$ are small and positive,
 $E(l)$ will have a shallow minimum at some value $l_{0}$,
similar to the behavior actually seen in
Ref. \onlinecite{LQ}. 

Furthermore, for droplets whose linear dimensions are similar to the 
minimum at $l_{0}$,
the apparent value of $\theta$ would appear to be close to
 zero. This may be the origin of the behavior in
 Refs. \onlinecite{PY, KM}, where simulations of Parisi's spin overlap
distribution function and the link overlap function were only consistent
 with the droplet picture if $\theta$ was taken to be about zero rather than
its domain wall value of order 0.20
in three dimensions \cite{3D} and 0.70 in four dimensions
\cite{4D}.   

To summarize: there exists a natural correction to the usual scaling 
formula for the energy of a droplet.
We have explicitly
verified its validity by numerical studies of large systems
 in two dimensions. If a
 similar correction is valid 
in three dimensions, it can explain a number of puzzling features which have
been seen in  simulations. A direct check of the validity of this
scaling correction needs numerical studies of larger three-dimensional
systems than those currently studied.

\begin{acknowledgments}
We would like to thank Alan Bray for many discussions.
 AKH obtained financial support from
the DFG (Deutsche Forschungsgemeinschaft) under grants Ha 3169/1-1 and
Zi 209/6-1. The simulations were performed at the ``Paderborn Center for
Parallel Computing'' and the ``Gesellschaft f|r Wissenschaftliche
Datenverarbeitung'' in Gvttingen, both in Germany.  
\end{acknowledgments}

\end{document}